\documentclass[aps,prl,twocolumn,superscriptaddress,floatfix]{revtex4}

\usepackage{graphicx}
\usepackage{dcolumn}
\usepackage{bm}
\usepackage{amsmath}
\usepackage{amssymb}
\usepackage{color}
\usepackage{mathtools}
\usepackage{amsfonts}

\begin{document}

\renewcommand{\vec}[1]{\mbox{\boldmath $#1$}}


\title{Observation of topological $Z_2$ vortex fluctuations \\
in the frustrated Heisenberg magnet NaCrO$_2$}


\author{K. Tomiyasu}
\email[Electronic address: ]{k.tomiyasu.sci@gmail.com}
\affiliation{Department of Physics, Tohoku University, Aoba, Sendai 980-8578, Japan}
\affiliation{Nissan ARC Ltd., Natsushima-cho 1, Yokosuka 237-0061, Japan}
\author{Y. P. Mizuta}
\affiliation{Graduate School of Engineering science, Osaka University, Toyonaka, Osaka 560-8531, Japan}
\author{M. Matsuura}
\affiliation{CROSS Neutron Science and Technology Center, IQBRC Bldg, Tokai, Ibaraki 319-1106, Japan}
\author{K. Aoyama}
\affiliation{Department of Earth and Space Science, Graduate School of Science, Osaka University, Osaka 560-0043, Japan}
\author{H. Kawamura}
\email[Electronic address: ]{h.kawamura.handai@gmail.com}
\affiliation{Molecular Photoscience Research Center, Kobe University, Kobe, 657-8501, Japan}


\date{\today}

\begin{abstract}
Spin fluctuations in the triangular-lattice Heisenberg antiferromagnet NaCrO$_2$ are investigated by means of quasi-elastic neutron scattering with high energy resolution and wide energy band. Two components with the following features are captured separately. They are pronounced at intermediate temperatures of 20-50 K. One with quite an extended lifetime corresponding to $\sim$$0.001E_{\rm ex}$ ($E_{\rm ex}$ the exchange energy) nearly disappears at low temperature 10 K, and the other with an extended lifetime $\sim$$0.01E_{\rm ex}$ survives there, identified as free $Z_2$ vortex and $Z_2$-vortex pair, respectively, in harmony with the $Z_2$ vortex theory.
\end{abstract}


\maketitle

\section{I. Introduction}
 Magnetic vortex, a topologically stable nano-scale spin structure object in an easy-plane magnet, has attracted much attention both from fundamental and application perspectives. The standard vortex is characterized by the winding number $n=0, \pm1, \pm2 \cdots$, the set corresponding to integers [$\mathbb{Z}$], where each $n$ forms distinct topological sector \cite{KT}. Some time ago, Kawamura and Miyashita theoretically predicted that the frustrated {\it isotropic\/} Heisenberg magnets in two dimensions (2D) could possess a different type of vortex characterized by the parity-like two-valued topological number [$\mathbb{Z}_2$] corresponding only to its presence/absence, a $Z_2$ vortex \cite{KM, Kawamura-review}. The non-collinear $SO(3)\equiv RP_3$ nature of the underlying spin correlation caused by frustration guarantees its topological stability via the relation $\Pi_1[SO(3)]=Z_2$.

 It was argued that the frustrated 2D Heisenberg magnets exhibit a topological transition at a finite temperature $T=T_V$ driven by the binding-unbinding of the $Z_2$ vortices \cite{KM, Kawamura-review, KawamuraYamamoto, KawamuraYamamoto2}. In sharp contrast to the Kosterlitz-Thouless transition \cite{KT}, the spin-correlation length remains finite even below $T_V$, and the low-$T$ state is a spin paramagnetic state with slow spin fluctuations characterized by the topologically broken ergodicity, dubbed as a ``spin-gel'' state \cite{KawamuraYamamoto, Kawamura-review}. Interestingly, the finite spin-correlation length makes the $Z_2$-vortex physics robust against weak perturbations like the weak three-dimensionality and the weak magnetic anisotropy {\it etc.\/}, which inevitably exist in real magnets.  Further, a recent theory showed that a free $Z_2$ vortex is capable of efficiently carrying spin density in connection with potential application of spintronics \cite{AoyamaKawamura}.

 However, the experimental observation of $Z_2$ vortex has been quite difficult and, in fact, remained elusive for years. In theory, the thermodynamic anomaly is extremely weak; for example, specific heat and magnetic susceptibility exhibit only a rounded non-singular maximum slightly above $T_V$, whereas an essential singularity occurring at $T=T_V$ is too weak probably to experimentally detect \cite{KM,Kawamura-review,KawamuraYamamoto}. Further, many experimental evidences, including specific heat \cite{Nakatsuji,Olariu}, magnetic susceptibility \cite{Nakatsuji,Olariu,Zhao,Nambu}, ESR \cite{Ajiro, Yamaguchi, Yamaguchi2, Hemmida, Gao}, NMR (NQR) \cite{Olariu, Takeya}, $\mu$SR \cite{Olariu, Yaouanc, Takeya, MacLaughlin, Zhao, Xiao}, neutron scattering \cite{Kadowaki, Nakatsuji, Nambu}, and optical measurements \cite{Kojima}, were accumulated for several materials but were all indirect in the sense of observing a free $Z_2$ vortex separately from a $Z_2$-vortex pair, which is the key ingredient to verify the $Z_2$ vortex theory.
 
 According to the theory, the $Z_2$ vortices dynamically emerge, with binding and unbinding, in the intermediate temperature range around $T_V$ \cite{KM, Kawamura-review, KawamuraYamamoto, KawamuraYamamoto2}. A free $Z_2$ vortex is generated only above $T_V$ via the unbinding of the $Z_2$-vortex pairs which exist both below and above $T_V$, and exhibits a diffusive motion with a long lifetime. Note that the free $Z_2$ vortex is a topologically protected object so that it can annihilate only by the collision with other free $Z_2$ vortices. Since there are few free $Z_2$ vortices just above $T_V$, the free-vortex lifetime is expected to be quite long there, while, well above $T_V$ where many free $Z_2$ vortices are generated, the mutual collision between them occurs frequently and their lifetime would become shorter. Since the $Z_2$-vortex-pair unbinding process is expected to occur most frequently around the specific-heat-peak temperature $T_{Cp}$, $T_{Cp}$ might give a measure of the upper-bound temperature for the presence of the long-lived isolated free $Z_2$ vortex. It was also simulated that a direct signature of free $Z_2$ vortex might be detectable as a quasi-elastic scattering in energy spectrum at $T$'s between $T_V$ and $T_{Cp}$~\cite{OkuboKawamura}. 

 Thus, in this study, we search for the spin fluctuations in the intermediate temperature range by means of quasi-elastic neutron scattering (QENS), known as the powerful method to observe spin fluctuations. We then find in the intermediate temperature range of 20-50 K two QENS components with extended lifetimes: One with quite an extended lifetime corresponding to $\sim 0.001E_{ex}$ ($E_{ex}$ the exchange energy) which disappears at low tempeartures $\lesssim 10$ K, and the other with an extended lifetime $\sim 0.01E_{ex}$ surviving there. The former component is ascribed to the free $Z_2$ vortex, while the latter one to the $Z_2$-vortex pairs. Our experimental data turn out to be consistent with the $Z_2$ vortex theory.

 The remaining part of the paper is organized as follows. In \S II, we explain the material and the experimental method employed in the present paper. Our experimental results are presented in \S III. The experimental data are interpreted in \S IV, where relation to the $Z_2$ vortex theory and to the earlier other experimental measurements is discussed. Finally, \S V is devoted to summary and concluding remarks. Some of the details of the experimental procedures are given in Appendix, i.e., [A] correction of the zero energy position, [B] fitting in elastic scattering scale, [C] examination of the dependence of the QENS signal on the integrated $|Q|$ range, and [D] additional information of the fitting parameters.

 



\section{II. Material and method} 

 NaCrO$_2$ is a promising candidate to realize the $Z_2$-vortex phenomena, as described below. This material provides a quasi-2D Heisenberg-spin (Cr$^{3+}$: $S=3/2$) system in rhombohedrally stacked triangular lattice with predominant nearest-neighbor inplane antiferromagnetic exchange coupling \cite{Hsieh,Hsieh2,Hsieh3}. Neither magnetic long-range order nor spin-glass transition is observed down to low temperatures. The specific heat exhibits a non-singular peak around $T_{Cp} \simeq 41$ K, with no other singular behavior \cite{Olariu}. The magnetic susceptibility also exhibits a broad maximum around 48 K. Neutron diffraction indicates that a spin arrangement is of short range less than 20 lattice spacings and of approximately 120-deg based structure \cite{Hsieh}. Further, NMR and $\mu$SR experiments indicate an unconventional gradual freezing or slowing down of spin fluctuations below $T_{Cp}$, which has been ascribed to the possible $Z_2$-vortex transition/crossover \cite{Olariu}. 

 QENS experiments were performed on the time-of-flight near-backscattering spectrometer, DNA, located at the beam port BL02, MLF, J-PARC spallation neutron source in Japan~\cite{DNA}. The high energy resolution of 0.004 meV and the energy range of $-0.03 < E < 0.10$ meV were obtained by using Si 111 analyzer and pulse shaping chopper rotating at 225 Hz with 3 cm slit (high-resolution; HR), whereas the wide energy range of $-0.5 < E < 1.5$ meV with moderate energy resolution of 0.015 meV was achieved without the pulse shaping chopper (wide band; WB).

 A powder sample of NaCrO$_2$ was synthesized by a solid-state reaction method, in which a stoichiometric mixture of Na$_2$CO$_3$ and Cr$_2$O$_3$ was ground, followed by the calcination at 900 $^{\circ}$C for 12 h in a blend-gas flow of 99{\%} Ar and 1{\%} H$_2$ with intermediate grinding and pelletizing. Approximately 4.1 g of the sample was placed onto an aluminum foil and shaped into a hollow cylinder with a thickness of 2 mm, a diameter of 12 mm, and a height of 50 mm in order to mitigate the neutron-absorption effects as much as possible. The cylinder was sealed in the thin aluminum container with He exchange gas that was placed under a cold head in a He closed-cycle refrigerator. The temperature $T$ was varied between 10 K and 50 K. Slight deviations of experimental zero energy position caused by experimental conditions are corrected as expalined in Appendix A.



%
%
\section{III. Experimental results}

\begin{figure}[htbp]
\begin{center}
\includegraphics[width=0.95\linewidth, keepaspectratio]{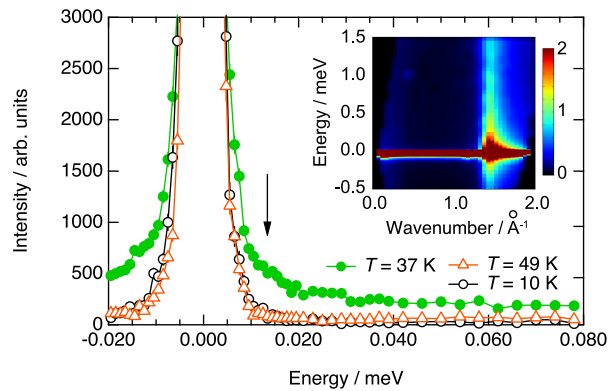}
\end{center}
\caption{\label{fig:raw} (Color online)
Typical measured neutron-scattering spectra. The inset shows the overall distribution in the $Q$ and $E$ space, obtained at 37 K in the WB mode. The main panel shows the HR energy spectra measured at three temperatures. The arrow indicates the existence of QENS. }
\end{figure}

 In this section, we present the results of our QENS experiment. First, we check the overall picture of intensity distribution in the wavenumber ($Q$) and energy ($E$) space in the WB mode, as shown in the inset of Fig.~\ref{fig:raw}. The QENS signal is confirmed in the $Q \approx 1.45$ {\AA}$^{-1}$. Further, the main panel of Fig.~\ref{fig:raw} shows the representative QENS energy spectra in the HR mode measured at three representative temperatures, extracted by integrating the $Q$ range from 1.3 to 1.6 {\AA}$^{-1}$. The QENS with 0.01-meV order of quite narrow energy width almost disappears at low-$T$ 10 K and high-$T$ 50 K and grows at medium-$T$ 37 K, as indicated by the arrow. 

 The fitting analysis of the QENS data is performed as follows. 
QENS intensity is generally described by
\begin{equation}
\label{eq:S_HR}
S(E) = \{ C_{\rm{el}} \delta(E) + B(E) \chi^{\prime \prime}(E) \} \otimes R(E) + C_0, 
\end{equation}
where $\delta(E)$ denotes the delta function for elastic incoherent scattering, $B(E)$ denotes the Bose population factor $[1-\exp \{ -E/(k_{\rm{B}}T) \}]^{-1}$, $\chi^{\prime \prime}(E)$ denotes the imaginary part of generalized magnetic susceptibility, $R(E)$ denotes the resolution function, for which we use the data measured for vanadium standard sample, $\otimes$ indicates the convolution integral by $R(E)$, and $C_0$ denotes the constant background. 

\begin{figure}[htbp]
\begin{center}
\includegraphics[width=0.95\linewidth, keepaspectratio]{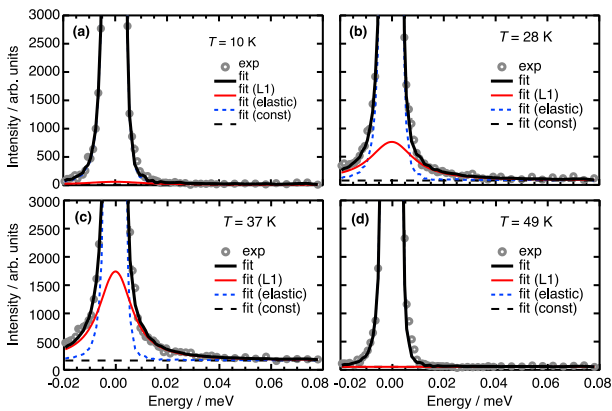}
\end{center}
\caption{\label{fig:spctr_HR} (Color online)
Fitting results for HR QENS data measured at several temperatures. The symbols indicate the observed data and the black solid curves indicate the calculated results. The red solid curves correspond to the QENS ($L_1$). The blue and black broken lines correspond to the elastic incoherent scattering and the constant background, respectively. (a) $T=10$ K, (b) $T=28$ K, (c) $T=37$ K, and (d) $T=49$ K. }
\end{figure}

For the HR data, we use the standard function,  
\begin{equation}
\label{eq:F_HR}
\chi^{\prime \prime}_{\rm{HR}}(E) = L_{1}(E) = C_{1} \frac{E\Gamma_1}{E^2 + \Gamma_1^2}, 
\end{equation}
where $\Gamma_1$ denotes the energy width in the fluctuation function. The typical fitting results are shown in Fig.~\ref{fig:spctr_HR}. Sufficiently good fitting is obtained for all the data. The fitting results of the HR data including the full elastic peak are also given in Fig.~5 of Appendix B, together with the residual errors. Sufficiently good fittings are also obtained in this scale, too.

\begin{figure}[htbp]
\begin{center}
\includegraphics[width=0.95\linewidth, keepaspectratio]{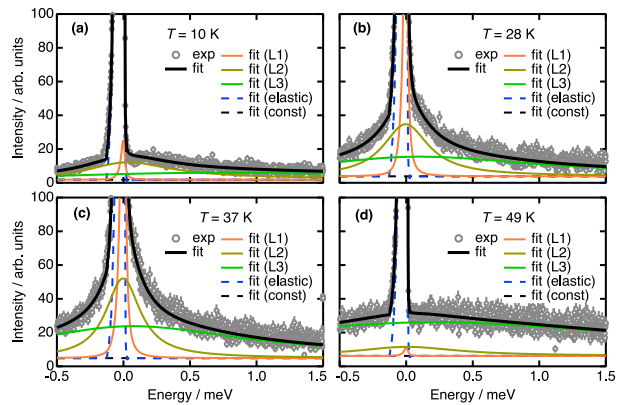}
\end{center}
\caption{\label{fig:spctr_WB} (Color online)
Fitting results for WB QENS data measured at several temperatures. The orange, gold, and green solid curves correspond to QENS. The other symbols, solid curves, and broken lines indicate the same as those defined in Fig.~\ref{fig:spctr_HR}.  }
\end{figure}

For the WB data, in addition to the $L_1$ component with 0.01-meV order width observed in the HR mode, another fluctuation component with 0.1-meV order width ($L_2$) is observed, as shown by gold lines in Fig.~\ref{fig:spctr_WB}. Furthermore, the meV-order component of spin-wave-like fluctuations was previously reported by higher-energy inelastic neutron scattering~\cite{Hsieh3} and the present WB data widely cover the energy range with a maximum of $E=1.5$ meV. Hence, the other meV-order component ($L_3$) is also necessary. 
To be precise, the spin-wave-like component is not QENS, but the $Q$-integrated energy spectrum exhibits a profile similar to QENS~\cite{Hsieh3}.
Thus, it should be reasonable to use the following model consisting of three components, 
\begin{equation}
\label{eq:F_WB}
\chi^{\prime \prime}_{\rm{WB}}(E) = \sum_{j=1}^{3}L_{j}(E) = \sum_{j=1}^{3}C_{j} \frac{E\Gamma_j}{E^2 + \Gamma_j^2}. 
\end{equation}

\begin{figure}[htbp]
\begin{center}
\includegraphics[width=0.95\linewidth, keepaspectratio]{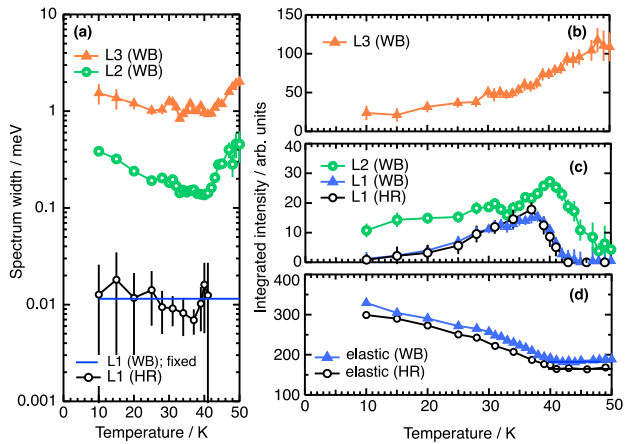}
\end{center}
\caption{\label{fig:prmtrs} (Color online)
Observed fitting parameters and their temperature dependences. (a) The spectrum widths $\Gamma_j$ ($j = 1$ to 3) are shown in a logarithmic scale. Integrated intensities of (b)(c) three fluctuation functions $L_j$ ($j = 1$ to 3), and (d) elastic components. The scales of arbitrary units are taken common with those of Figs.1 and 2 for HR, and of Fig. 3 for WB, while $L_1$ (HR) data in (c) and elastic (HR) data in (d) are multiplied by 0.54 and 0.44, respectively, for illustration.
}
\end{figure}

However, the 0.01-meV order $\Gamma_1$ is too small to re-determine by fitting the WB data. Therefore, it is better to use the values already obtained by fitting the HR data. Further, they exhibit no significant temperature dependence within the error bars, though somewhat rough, as shown in Fig.~\ref{fig:prmtrs}(a) later, Thus, we fix $\Gamma_1$ to the HR mean value, 0.012 meV. The typical fitting obtained is shown in Fig.~\ref{fig:spctr_WB}. Sufficiently good fitting is obtained for all the data. The fitting results of the WB data including the full elastic peak are given in Fig.~6 of Appendix B, together with the residual errors. Sufficiently good fittings are also obtained.

 In Figs.~1 and 2, we integrated the intensity of QENS for the relatively wide $Q$-region of $1.3<|{\bm Q}|<1.6$ \AA$^{-1}$ to capture all magnetic signal at all temperatures even if the magnetic wavevector changes. To examine how the results depend on the integrated $|{\bm Q}|$ range, we also try similar data analysis with a narrower $|{\bm Q}|$ range of $1.4<|{\bm Q}|<1.5$ \AA$^{-1}$ both for the HR and WB data, and the results are given in Appendix C. As shown in its Fig.~7, narrow QENS signals quite similar to the ones of Figs.~1 and 2 are obtained in the intermediate temperature range, demonstrating that our conclusion does not change with the $Q$-width of integrated region.

 We summarize the obtained spectrum widths and integrated intensities in Fig.~\ref{fig:prmtrs}. Figure~\ref{fig:prmtrs}(a) shows the three widths in a logarithmic scale, 0.01-meV order $\Gamma_1$, 0.1-meV order $\Gamma_2$, and 1-meV order $\Gamma_3$. The scale of the smallest $\Gamma_1$ corresponds to surprisingly slow spin fluctuations of 10$^{-3}$ times the magnitude of exchange energy $E_{\rm ex} \approx 9$ meV and thermal energy 40 K $\approx$ 4 meV, where $E_{\rm ex}$ is estimated as $J_{1}S(S+1)$, nearest-neighbor antiferromagnetic exchange $J_{1}=2.4$ meV, and spin $S=3/2$; these values were reported by Curie-Weiss fitting in magnetic susceptibility and inelastic neutron scattering study of spin waves \cite{Hsieh2}. Further, $\Gamma_1$ exhibits no significant temperature dependence, as mentioned above. As the temperature decreases, $\Gamma_2$ decreases rapidly, bottoms around 40 K, and then increases slightly. The temperature dependence of $\Gamma_3$ is similar to that of $\Gamma_2$ but the degree of change is relatively small. 

Figures~\ref{fig:prmtrs}(b) to \ref{fig:prmtrs}(d) show the temperature dependence of integrated intensities of fluctuation and elastic components. As the temperature decreases from 50 K, the $L_3$ fluctuation component with largest integrated intensity in $L_1$, $L_2$, and $L_3$ monotonically decreases [Fig.~\ref{fig:prmtrs}(b)] and, instead, the elastic component (nearly static and/or static) monotonically increases below 41 K [Fig.~\ref{fig:prmtrs}(d)], as normally expected for the critical slowing down of dynamics. 
In fact, the onset temperature, at which the nearly static component begins to rise, exhibits the systematic tendency with other experimental data in accordance with the energy resolution. Neutron diffraction (over meV order of low resolution) exhibits the onset temperature of approximately 45 K, present high-resolution elastic scattering (1 to 10-$\mu$eV order) 41 K, and NMR and $\mu$SR ($\mu$eV to neV) approximately 30 K~\cite{Olariu}. That is, as the energy resolution becomes finer (time scale becomes slower), the onset temperature decreases. 

In contrast, the behaviors of two $L_1$ and $L_2$ fluctuation components are nonmonotonic [Fig.~\ref{fig:prmtrs}(c)]. As the temperature decreases from 50 K, the $L_2$ intensity increases below 48 K, exhibits the maximum around 40 K, then, monotonically decreases, but survives even at 10 K. The $L_1$ intensity increases below 43 K, exhibits the maximum around 37 K, then, monotonically decreases, and almost disappears at 10 K. 
We also remark that the HR and WB temperature dependences of $L_1$ intensities are plotted overlaid, which are consistent with each other, supporting the validity of the fitting results. For completeness, we also show the temperature dependence of the constant background in Fig.~8 in Appendix D.

 Thus, these QENS experiments succeed in capturing the characteristic spin fluctuations ($L_1$ and $L_2$ components), which are orders of magnitude slower than the energy scale of exchange coupling and thermal fluctuations, and grow only in the critical temperature range between 20 and 50 K. In particular, the $L_1$ component is most enhanced around 37 K, which is clearly lower than specific-heat $T_{Cp}\simeq 41$ K. These features are different from normal ordering and critical scattering.

\section{IV. Interpretation and discussion}

 The most probable theory to describe the spin fluctuations in NaCrO$_2$ is the $Z_2$ vortex model \cite{KM, Kawamura-review, KawamuraYamamoto, KawamuraYamamoto2}, as discussed with the circumstantial evidences in \S I and \S II. Thus, we check the observed four components ($L_1$, $L_2$, $L_3$, and elastic) against this model. 

 First, this model also consists of mainly four components, (1) free $Z_2$-vortex fluctuations, (2) $Z_2$-vortex-pair fluctuations, (3) spin-wave-like fluctuations including paramagnons, and (4) spin gel described by the essentially dynamical but nearly static short-range order with 120-deg based structure. Nearly static spin gel most likely corresponds to the elastic component. Spin-wave-like fluctuations also most probably correspond to the $L_3$ component, which was originally set with those in mind in Eq.~(\ref{eq:F_WB}). The overall temperature dependence, in which spin-wave-like fluctuations decrease [Fig.~\ref{fig:prmtrs}(b)] and elastic spin gel instead grows on decreasing the temperature [Fig.~\ref{fig:prmtrs}(d)], is also natural both theoretically and experimentally. 

 The temperature dependence of the spectral width $\Gamma_3$ shown in Fig.~4(a), i.e., the spin-wave $L_3$ component including paramagnons, can be understood as follows. In general, as the temperature decreases towards anomaly temperature in the high-temperature region, the spectral width of paramagnons narrows due to the suppression of thermal fluctuations. In the low-temperature region below anomaly temperature, where the well-defined spin-wave dispersion forms, the dispersion becomes a little steeper, and the pseudo-DOS spectral distribution extends a little to the higher energy side. Further, frustration causes the temperature range of these critical phenomena to expand. The spin-wave dispersion in NaCrO$_2$ becomes observable below around 40 K in inelastic neutron scattering \cite{Hsieh2}. Thus, all of these are consistent with the temperature dependence of $\Gamma_3$.

 Next, we discuss whether the components (1) and (2) of the $Z_2$ vortex theory, i.e., the free $Z_2$ vortex and the $Z_2$ vortex pair, can be identified to the $L_1$ and $L_2$ components of the experiment. Theoretically, the $Z_2$ vortex system is described by a topological number of $Z_2$ vorticity, $z = 1$ and 0, where $z=1$ denotes the free $Z_2$ vortex and $z=0$ denotes the $Z_2$-vortex pair or no vortex. Ideally, only $z=0$ is allowed below $T_V$, whereas both $z=0$ and 1 are allowed above $T_V$. Interestingly, this kind of distinction is observed, as shown in Fig.~\ref{fig:prmtrs}(c). At the low temperature $T=10$ K, the $L_1$ component almost disappears and the $L_2$ component finitely survives. At higher temperatures, both the components exist. Thus, the $L_1$ and $L_2$ components can be identified to the free $Z_2$ vortex ($z=1$) and the $Z_2$-vortex pair ($z=0$), respectively. 

 The earlier $\mu$SR study revealed that $T_1^{-1 }$ exhibits a broad peak centered around 25-30 K, suggesting that low-energy spin fluctuations appear there, and survives at 10 K \cite{Olariu}. The $T_1^{-1}$ is accompanied by the stretched exponent $\alpha = 0.5$, indicating a nontrivial distribution of $T_1^{-1}$ or of muon sites, which makes the interpretation of $T_1^{-1}$ difficult. The observation of the non-vanishing $T_1^{-1}$ even in the lower temperature range below around 10 K might come from the contribution of such broad distribution of the relaxation time.

 In the present QENS study, on the other hand, by using the DNA spectrometer with high energy resolution and with a wide energy range of three orders of magnitude from 0.003 meV to 1 meV \cite{DNA}, we have succeeded in directly observing the energy-resolved spin fluctuations over a wide energy range with the same spectrometer and under the same $Q$ range and experimental conditions, allowing us to capture the overall and more resolved picture of spin fluctuations in comparison with the $Z_2$ vortex theory for the first time. In this way, we have succeeded in decomposing low-energy spin fluctuations into the two components, $L_1$, which vanishes at 10 K, and $L_2$, which survives there.

 Of course, the type of the expected anomalies might differ between the QENS and the $\mu$SR. Taking an example of simple ferromagnets, an inflection point of the temperature vs. magnetization (the order parameter of ferromagnets) curve can be employed as an experimental indicator of $T_c$, whereas the diverging point of the temperature vs. the susceptibility curve can also be employed for that purpose. In our present case, the $L_1$-peak intensity of QENS represents the number of free $Z_2$ vortices serving as the order parameter (or rather the disorder parameter) of the $Z_2$-vortex ordering, while the $T_1^{-1}$ of $\mu$SR corresponds to the dynamical susceptibility. Hence, one may expect that the peak temperature of the $\mu$SR (or ESR) $T_1^{-1}$, which was reported to be $25\pm 5$ K \cite{Olariu, Hemmida}, might roughly correspond to the point at which the QENS $L_1$-peak intensity exhibits a sharp drop on decreasing $T$. Although it is difficult to identify a truly sharp anomaly in our data of the $L_1$ intensity, it appreciably decreases across 25 K on decreasing $T$, consistently with such an expectation. Of course, anomaly temperatures observed by different experimental probes with different energy scales do not necessarily agree, and such a correspondence should be regarded as rough correspondence. For more precise estimation and deeper understanding of the vortex transition temperature, a future systematic QENS study for other $A$CrO$_2$ ($A$ = H, Li, Na) compounds might be helpful.

 Thanks to the $Z_2$ topological protection, an isolated free $Z_2$ vortex is expected to be long-lived compared to the energy scale of the exchange coupling and thermal fluctuations, being consistent with the observation of narrow width of the $L_1$ component, $\Gamma_1\sim \tau_1^{-1}$. However, the observed $\Gamma_1 \sim 0.001E_{\rm ex}$ is even narrower than the value expected from the previous spin-dynamics simulation $\sim 0.01E_{\rm ex}$ \cite{OkuboKawamura}. In order to examine whether this amount of narrow width (long lifetime) is possible in the $Z_2$ vortex scenario, we have recently extended the earlier simulation of Ref.~\cite{OkuboKawamura} to higher energy resolution (longer simulation time) and to larger lattices, to find that the width of $\sim 0.001E_{\rm ex}$ is indeed possible \cite{Mizuta}. By contrast, a $Z_2$-vortex pair consists of two nearby $Z_2$ vortices that can destroy each other, being consistent with the observation of relatively large width (short lifetime) $\Gamma_2\sim \tau_2^{-1}$ of the $L_2$ component, that is, the magnitude relationship of $\Gamma_1 < \Gamma_2$ $(\tau_1 > \tau_2)$.



However, it may be only in the low-temperature range that the $L_1$ and  $L_2$ components can be clearly decomposed as free and paired $Z_2$ vortices, respectively. 
In simulations~\cite{KM,Kawamura-review,KawamuraYamamoto}, at $T\gtrsim T_{Cp}$, the high-density free $Z_2$ vortices spatially mingle with the $Z_2$-vortex pairs with mutual collisions, and their lifetime would be shorter and comparable to each other. Hence, around and above $T_{Cp} \sim 41$ K, the $L_2$ component is considered the mixture of $Z_2$-vortex pairs and short-lived free $Z_2$ vortices.

Further, in the higher-temperature range, as the $Z_2$ vorticity itself becomes ill-defined due to the loss of the minimum amount of short-range order, and the $Z_2$-vortex mixture would be indistinguishable from other high-energy excitations. This might be the reason behind our observation that the $L_2$ component goes away and only the $L_3$ component remains above around 50 K.

 As argued above, although it is rather difficult to precisely estimate $T_V$ from the present QENS data, our data are consistent with the earlier $\mu$SR and ESR estimates of $25\pm5$ K \cite{Olariu, Hemmida}. We could also estimate $T_V$ from the $Z_2$ vortex theory giving the relation, $T_V = 0.285E_{\rm ex} = 0.285J_{1}S(S+1)$ \cite{KM, KawamuraYamamoto}. Using $J_1 = 2.4$ meV = 28 K, estimated by the meV-order inelastic neutron scattering and spin wave analysis \cite{Hsieh2}, we obtain $T_V = 0.285 \cdot 28 \cdot (3/2)(3/2+1) = 30$ K. Using another value $J_1 = 20$ K \cite{Olariu}, we obtain $T_V = 0.285 \cdot 20 \cdot (3/2)(3/2+1) = 21$ K. Thus, all the studies seem to suggest the $T_V$-value in the range of $25\pm5$ K.

 Although the free $Z_2$ vortices disappear below $T_V$ in theory, the corresponding $L_1$ intensity exhibits a low-$T$ tail below 25 K before vanishing around 10 K [Fig.~\ref{fig:prmtrs}(c)]. This rounding-off behavior might arise from either of the following two effects, i.e., i) the non-ideal effects in experiments, e.g., the impurities or defects might pin the free $Z_2$ vortex at low $T$, slow relaxation process  in the spin-gel state below $T_V$ might hamper the efficient pair annihilation of the far-apart free $Z_2$ vortices, or ii) the powder nature of the sample tends to make the $L_1$ peak of QENS ``normalized'', consisting of the contributions not purely from the free $Z_2$ vortex, but also from the loosely bound $Z_2$-vortex pair convoluted with.

 We wish to discuss the point ii) further in connection with the recent theoretical calculation \cite{Mizuta}. Ref.\cite{Mizuta} suggests that the width and the intensity of the QENS peak vary with the wavevector $\bm{Q}$, becoming narrower as $\bm{Q}$ approaches the $K$ point, eventually being as narrow as $\sim 0.001J$ which is of the same order as the $L_1$-peak width. Such a narrow peak arising from $\bm{Q}$'s close to the $K$ point has then been ascribed to the free $Z_2$ vortex, while a bit wider peak arising from $\bm{Q}$'s slightly away from the $K$ point with the loosely bound $Z_2$-vortex pair.

 Since our sample is the powder, the $L_1$ peak of QENS might contain not only the contributions from $\bm{Q}$'s very close to the $K$ point, but also contain those slightly away from the $K$ point. Yet, theory has indicated that near $T_V$ the QENS intensity is dominated by the contributions from $\bm{Q}$'s very close to the $K$ point associated with the free $Z_2$ vortex (see Fig.~1c of Ref.\cite{Mizuta}), and consequently, a sharp QENS peak of its width $0.001J$ would be visible even after the powder average, consistently with the present QENS observation.  

 In this way, the powder-averaged $L_1$-peak intensity might not be a contribution purely from the free $Z_2$ vortex, but rather be a normalized one convoluted with the contribution from loosely bound $Z_2$-vortex pair. This raises an interesting possibility also for the temperature ($T$) dependence of the experimental $L_1$-peak width. As $T$ is decreased toward $T_V$ from above, the number of both free $Z_2$ vortices and $Z_2$-vortex pairs would decrease, but with different rates, the former being decreased more strongly than the latter simply because only the former should vanish at $T_V$. While the peak width generally decreases with decreasing $T$ at a given fixed $\bm{Q}$, the aforementioned $T$-variation of the relative weights of different $\bm{Q}$'s would weaken the decrease of the peak width, which seems consistent with the experimental observation that the $T$-dependence of the $L_1$-peak width is rather weak.

 From a theoretical perspective, the $Z_2$-vortex physics is basically of topological nature,  rather insensitive to many details of the system. For example, the nature of the $Z_2$-vortex order is expected to be insensitive to whether the underlying magnetic short-range order is either commensurate or incommensurate so long as it is noncollinear. Especially, since the spin correlation length remains finite even below $T_V$, the $Z_2$-vortex order is robust against the weak perturbative interactions, e.g., the weak interplane coupling. In the case of our target magnet NaCrO$_2$, neutron-scattering measurements by Hsieh et al indicated that the in-plane translation vector became incommensurate below 30 K due to the effect of the weak interplane coupling \cite{Hsieh}. The same measurements also indicated that the 3D magnetic order did not arise even below 30 K, the interplane correlation length staying only about four layers. This means that the system remains essentially 2D even below 30 K. Note that, concerning the ordering process or phase transition where the long-distance behavior is crucially important, four-layers system should still be regarded as 2D system. Hence, it is theoretically expected that the same $Z_2$-vortex physics applies to NaCrO$_2$, the only possible change being a deformation of the short-scale spin structure caused by the interplane interaction, which, however, maintains the $SO(3)$ topology, a crucial ingredient of the $Z_2$-vortex physics.

\section{V. Summary and concluding remarks}

 In summary, by using the single spectrometer with high energy resolution and wide energy band, we succeeded in clarifying the four components that construct the $Z_2$ vortex theory prediction for the first time, {\it i.e.\/}, spin gel, free $Z_2$ vortex, $Z_2$-vortex pair, and spin wave. The $Z_2$ vortex theory can naturally explain many experimentally observed characteristics, which verifies the $Z_2$ vortex theory and its real materiality.

 The $Z_2$ vortex, although predicted long time ago \cite{KM}, has got relatively little attention until now. However, its new importance in terms of spintronics was also demonstrated by the recent theory, in which this topologically protected spin texture can be a stable and efficient carrier of spin density~\cite{AoyamaKawamura}. Hopefully, the present result may promote further studies on the thermodynamic and transport properties of the $Z_2$ vortex.

\acknowledgments
We thank Dr. Y. Kousaka for supporting the sample preparation. The neutron experiments were performed with the approval of J-PARC (2019A0295). This study was financially supported by MEXT and JSPS KAKENHI (JP17H06137, JP18K03503). 


\appendix

\section{A. Fitting of zero energy}
In order to incorporate the slight deviations of experimental zero energy position $E_0$ caused by subtle differences in experimental conditions, for example, accelerator conditions,  chopper speed fluctuations, and sample positions, we fit the $E_0$ value in each spectrum under the constraint that all the $E_0$'s of delta function, $L_1$, $L_2$, and $L_3$ are identical. 
Precisely, QENS intensity is described by
\begin{equation}
\begin{split}
\label{eq:S_HR}
S(E) &= \{ C_{\rm{el}} \delta(E^{\prime}) + B(E^{\prime}) \chi^{\prime \prime}(E^{\prime}) \} \otimes R(E^{\prime}) + C_0 \\
E^{\prime} &= E-E_0. 
\end{split}
\end{equation}
In all the HR data, the obtained $E_0$ values are approximately only $4\times10^{-4}$ meV and the difference among the spectra measured at different temperatures are only $3\times10^{-5}$ meV, which are negligible for both the typical $\Gamma_1$, $1\times10^{-2}$ meV, and the resolution, $4\times10^{-3}$ meV.

\section{B. Fitting in elastic scattering scale}

\begin{figure}[htbp]
\begin{center}
\includegraphics[width=0.95\linewidth, keepaspectratio]{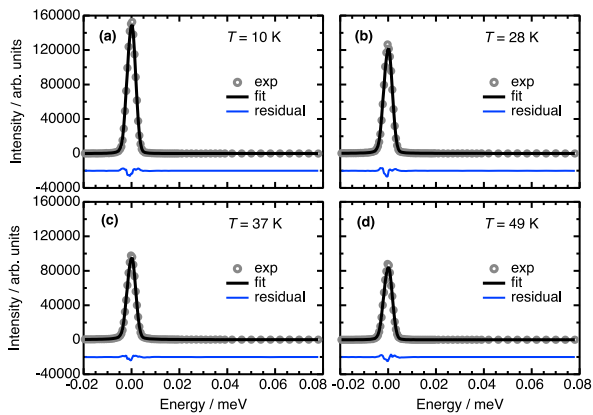}
\end{center}
\caption{\label{fig:SM_1} HR data of the full elastic peak with residual errors. The symbols indicate the observed data. The black solid curves indicate the calculated results. The blue lines correspond to the residual errors. (a) $T=10$ K, (b) $T=28$ K, (c) $T=37$ K, and (d) $T=49$ K. The scales of arbitrary units are taken common as those of Figs.~1 and 2 in the main text.
} 
\end{figure}

\begin{figure}[htbp]
\begin{center}
\includegraphics[width=0.95\linewidth, keepaspectratio]{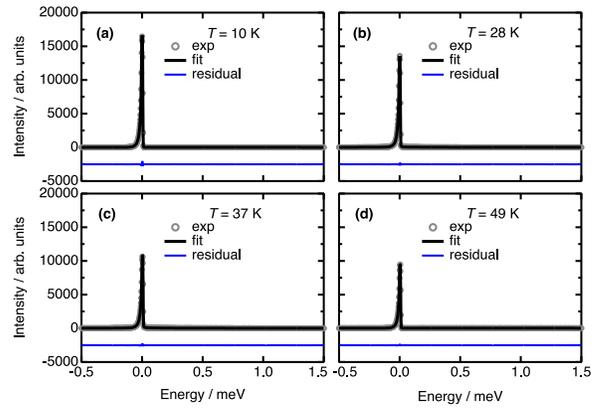}
\end{center}
\caption{\label{fig:SM_2} WB data of the full elastic peak with residual errors. The symbols indicate the observed data. The black solid curves indicate the calculated results. The blue lines correspond to the residual errors. (a) $T=10$ K, (b) $T=28$ K, (c) $T=37$ K, and (d) $T=49$ K. The scales of arbitrary units are taken common as those of Fig.~3 in the main text. 
} 
\end{figure}

 Figs.~\ref{fig:SM_1} and \ref{fig:SM_2} show the HR and WB data of full elastic peak with residual errors, respectively. Sufficiently good fittings are obtained in this scale, too. 

The QENS signal is very weak compared to the elastic scattering component, but the instrument used for the measurements (DNA) exhibits the world's highest signal-to-noise ratio (very low background) among similar instruments \cite{DNA}. The signal-to-noise ratio of DNA reaches 100,000, which is nearly 100 times better than others, that is, the best in the world. This great performance allows for the fine fitting. 

The tail from elastic peak on the negative energy side in Fig.~6 is due to tail of neutron pulse for the wide-band mode, whereas the larger tail on the positive energy side in Fig.~3 in the main text is due to QENS enhanced by the Bose factor.


\section{C. Examination of the dependence of the QENS signal on the integrated $|\bm{Q}|$ range}

In the main text, we integrate the intensity of QENS for the relatively wide $Q$-region of $1.3<|{\bm Q}|<1.6$ \AA$^{-1}$ to capture all magnetic signal at all temperatures even if the magnetic wavevector changes. To examine how the results presented in the main text depend on the integrated $|{\bm Q}|$ range, we also try similar data analysis with a narrower $|{\bm Q}|$ range of $1.4<|{\bm Q}|<1.5$ \AA$^{-1}$, and the results corresponding to Fig.~1 in the main text are shown in Fig.~7. Intensities of Vanadium data in these plots are normalized to the $T$=10K data at $E$=0. In the high-resolution (HR) data, narrow QENS with 0.01 meV order ($L_1$) appears in the intermediate temperature range, which is consistent with Fig.~1. For the wide-band (WB) data, QENS with 0.1 meV order ($L_2$) also appears in the intermediate temperature range. Thus, our conclusion does not change with the $Q$-width of integrated region.

\begin{figure}[htbp]
\begin{center}
\includegraphics[width=0.95\linewidth, keepaspectratio]{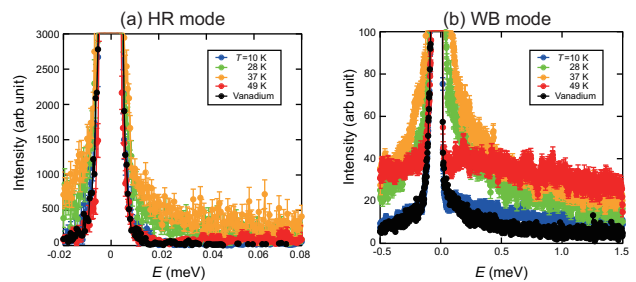}
\end{center}
\caption{\label{fig:Qrange}Neutron-scattering spectra corresponding to the narrower $|\bm{Q}|$ range from 1.4 to 1.5 \AA$^{-1}$ measured at four temperatures, (a) in the HR mode, and (b) in the WB mode.
}
\end{figure}

\section{D. The fitting parameters}

 In Fig.~4 in the main text, we have shown the temperature dependence of the width of the $L_1 - L_3$ peaks as well as the integrated intensity of the $L_1- L_3$ peaks and the elastic component. In this section of Appendix, we show for completeness the corresponding quantity for the constant  background. Fig.~8 shows the temperature dependence of the background parameters in the HR and WB modes. In the HR mode, the fitted background parameters exhibit a similar temperature dependence to the $L_2$ component. This is quite natural result, because the $L_2$ component identified in the WB mode is captured in the HR mode as constant background due to the narrow energy range of the HR mode.

 In the WB mode, a phononic background linearly proportional to the temperature should be considered, in addition to the temperature-independent instrumental background. We obtained such a background $a + b'T$ first at $Q=0.8$ {\AA}$^{-1}$ where no magnetic signal exists. Since the phononic background is proportional to $Q^2$, the background parameter $b$ at $Q_{mag}=1.45$ {\AA}$^{-1}$ was estimated  by multiplying $b'$ by the square of the ratio of $Q_{mag}=1.45$ {\AA}$^{-1}$ to 0.8 {\AA}$^{-1}$, and the results are shown in Fig.~8.

\begin{figure}[htbp]
\begin{center}
\includegraphics[width=0.95\linewidth, keepaspectratio]{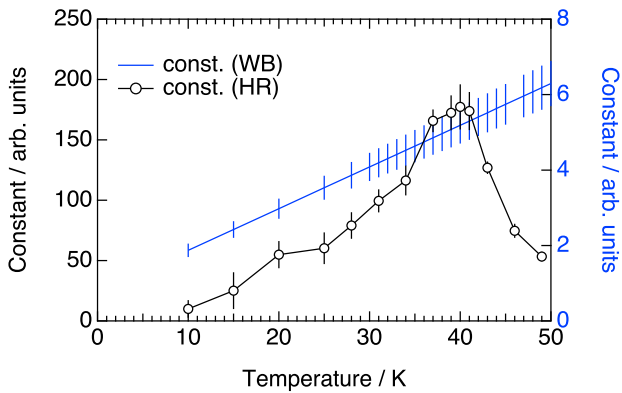}
\end{center}
\caption{The temperature dependence of the constant background parameters. The scales of arbitrary units are taken common with those of Figs.~1 and 2 in the main text for the HR mode, and of Fig.~3 in the main text for the WB mode.
} 
\end{figure}

\end{document}